# Design of Lead-free Inorganic Halide Perovskites for Solar Cells via Cation-transmutation

Xin-Gang Zhao[1,‡], Ji-Hui Yang[2,‡], Yuhao Fu[1], Dongwen Yang[1], Qiaoling Xu[1], Liping Yu[3], Su-Huai Wei[4,*] and Lijun Zhang[1,*]

[1]Department of Materials Science and Engineering and Key Laboratory of Automobile Materials of MOE and State Key Laboratory of Superhard Materials, Jilin University, Changchun 130012, China

Jilin University, Changchun 130012, China

[2]Department of Materials Science and Nanoengineering, Rice University, Houston, Texas 77005, USA

[3]Department of Physics, Temple University, Philadelphia, PA 19122, USA

[4]Beijing Computational Science Research Center, Beijing 100094, China

**Abstract:** Hybrid organic-inorganic halide perovskites with the prototype material of $CH_3NH_3PbI_3$ have recently attracted intense interest as low-cost and high-performance photovoltaic absorbers. Despite the high power conversion efficiency exceeding 20% achieved by their solar cells, two key issues — the poor device stabilities associated with their intrinsic material instability and the toxicity due to water soluble $Pb^{2+}$ — need to be resolved before large-scale commercialization. Here, we address these issues by exploiting the strategy of cation-transmutation to design stable inorganic Pb-free halide perovskites for solar cells. The idea is to convert two divalent $Pb^{2+}$ ions into one monovalent $M^+$ and one trivalent $M^{3+}$ ions, forming a rich class of quaternary halides in double-perovskite structure. We find through first-principles calculations this class of materials have good phase stability against decomposition and wide-range tunable optoelectronic properties. With photovoltaic-functionality-directed materials screening, we identify eleven optimal materials with intrinsic thermodynamic stability, suitable band gaps, small carrier effective masses, and low excitons binding energies as promising candidates to replace Pb-based photovoltaic absorbers in perovskite solar cells. The chemical trends of phase stabilities and electronic properties are also established for this class of materials, offering useful guidance for the development of perovskite solar cells fabricated with them.

## 1. Introduction

Hybrid organic-inorganic halide perovskites with a chemical formula of $AM^{IV}X^{VII}_3$, where A represents a small monovalent organic molecule, $M^{IV}$ is a divalent group-IVA cation and $X^{VII}$ is a halogen anion, have recently attracted a tremendous amount of attention in the photovoltaic community.[1–15] Current record power conversion efficiency (PCE) of solar cells based on them has been boosted from initial value of 3.8%,[8] step by step,[16,17,5,18–21] to current 22.1%.[22] During this process the breakthrough step is the first solid-state perovskite solar cell designed by Kim *et al.* yielding the PCE exceeding 9%,[16] which triggered significant perovskite solar cell research activities and made PCEs dramatically enhanced within only shortly seven years. Such a rapid progress far surpasses the cases of many conventional thin film solar cells (*i.e.*, fabricated with crystalline Si, CdTe, Cu(In,Ga)Se$_2$, etc.) that achieved similar PCEs after decades of efforts. The high PCE of halide perovskites originates from their correlative intrinsic material properties, including suitable band gaps and high threshold light absorption,[23,24] defect-tolerant feature,[25–29] ultra-long carrier diffusion length,[9] low exciton binding energy,[30,31] balanced electron and hole mobility,[32,33] etc. These unique properties, accompanying with the low-cost solution-based fabrication routes, make them ideal candidates as new-generation photovoltaic absorbers.

Despite enormous success of $AM^{IV}X^{VII}_3$ perovskites in solar cell applications, challenges are still standing in their way to large-scale commercial applications. The first serious issue is their poor long-term device stability, especially under heat and humidity conditions, which could be partially attributed to the intrinsic thermodynamic instability of $AM^{IV}X^{VII}_3$.[34–36] While the underlying mechanism remain unclear, it is likely because of the organic cations involved that correspond to rather loose chemical bonding, and their inherent instability.[3,37,38] Experimentally, it has been demonstrated that mixing a small amount of inorganic cations such as $Cs^+$ with methylammonium ($CH_3NH_3^+$, *i.e.*, $MA^+$)/formamidinium ($CH_3(NH_2)_2^+$, *i.e.*, $FA^+$) at the A site significantly increases the stabilities of perovskite films.[2,39–41] Contrast to the organic-inorganic hybrid perovskites, purely inorganic $CsPbX^{VII}_3$ perovskites exhibit excellent thermal stability.[3,38] Secondly, because the upper group-IVA elements such as Sn and Ge at the $M^{IV}$ site tend to be oxidized from divalent $Sn^{2+}/Ge^{2+}$ to tetravalent $Sn^{4+}/Ge^{4+}$,[32,42–44] and thus cause the instability issue, current halide perovskite based solar cells with high PCEs exclusively contain the toxic element — Pb. This will inevitably cause potential environmental concerns for large-scale solar cell devices.[45–47] Consequently, it is of great interest to find alternative halide perovskites consisting of completely inorganic components having good stability and meanwhile made of less toxic elements.

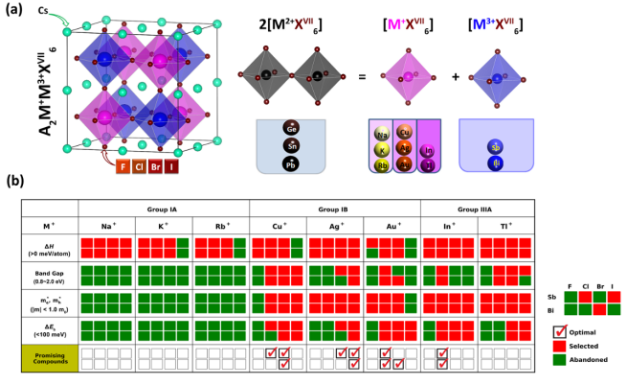

**Figure 1.** (a) Space of candidate $A_2M^+M^{3+}X^{VII}_6$ perovskites for materials screening: left panel shows adopted double-perovskite structure, and right panel shows schematic idea of atomic transmutation. (b) Materials screening process by considering gradually the properties relevant to photovoltaic performance, i.e., decomposition enthalpy ($\Delta H$), band gap, carriers effective masses ($m_e^*, m_h^*$), and exciton binding energy ($\Delta E_b$). The red squares mean the materials passing the screening (Selected) and the green ones mean not passing (Abandoned). The optimal non-toxic $A_2M^+M^{3+}X^{VII}_6$ perovskites satisfying all the criterions are marked with red checks.

To eliminate the toxic Pb, while the straightforward idea is considering other divalent cations beyond group-IVA elements, it turns out that the choice is limited and the resulting compounds have poor optoelectronic properties for solar cells (e.g., too large band gaps and heavy carrier effective masses).[48–52] Alternatively, one can consider to transmute two divalent $Pb^{2+}$ ions into one monovalent ion $M^+$ and one trivalent ion $M^{3+}$, i.e., $2Pb^{2+} \rightarrow M^+ + M^{3+}$, by keeping the total number of valance electrons unchanged at $M^{IV}$ sites. Known as atomic transmutation, this design strategy has led to great success in finding new materials with improved functionalities such as solar absorbers. Typical example is the evolution of binary ZnSe to ternary $CuGaSe_2$ and then to quaternary $Cu_2ZnSnSe_4$: ZnSe has a wide band gap of 2.8 eV, which is too large for solar cell application. By transmuting two $Zn^{2+}$ ions into one $Cu^+$ and one $Ga^{3+}$, i.e., $2Zn^{2+}\rightarrow Cu^+ + Ga^{3+}$, $CuGaSe_2$ is obtained with a band gap of 1.7 eV. Further atomic transmutation, i.e., $2Ga^{3+}\rightarrow Zn^{2+}+Sn^{4+}$, leads to $Cu_2ZnSnSe_4$ with a band gap of 1.0 eV.[53,54] Both $CuGaSe_2$ and $Cu_2ZnSnSe_4$ based compounds [i.e., $Cu(In,Ga)Se_2$-CIGS and $Cu_2ZnSnS_4$-CZTS] have been used in solar cells showing high PCEs of above 20% and 10%, respectively. With this transmutation strategy applied, a novel class of Pb-free quaternary materials in the formula of $A_2M^+M^{3+}X^{VII}_6$ with the double-perovskite structure may form. Considering that the $6s^2p^0$ configuration of $Pb^{2+}$ is believed to be responsible for the unique optoelectronic properties of $AM^{IV}X^{VII}_3$ perovskites,[24,55] available choice of $M^{3+}$ can be isoelectronic $Bi^{3+}$ and $Sb^{3+}$ and $M^+$ can be any size-matching monovalent cations. Compatible with the smaller sizes of $Bi^{3+}/Sb^{3+}$ than $Pb^{2+}$, the small inorganic cations (rather than the large organic cations commonly used in $AM^{IV}X^{VII}_3$) can be adopted at the A site to stabilize the perovskite lattice, opening the avenue for achieving the inorganic halide perovskites with good stabilities. In fact, several such double-perovskite compounds $A_2M^+M^{3+}X^{VII}_6$ have been synthesized since 1970s[56–60] but never considered for photovoltaic applications. Only until quite recently proposals of using $Cs_2AgBiCl_6$ and $Cs_2AgBiBr_6$ as potential solar absorbers were put forward.[61–63] Though exhibiting good stability when exposed to air, neither of them show superior photovoltaic performance due to their indirect band-gap feature and large gap values (above 2 eV).[61,63,64] Given the fact that the group of quaternary $A_2M^+M^{3+}X^{VII}_6$ perovskites is much broader owing to its multinary nature, one wonders if other members may have advantages. In such context, a systematic exploration of $A_2M^+M^{3+}X^{VII}_6$ perovskites by considering as many as possible (A, $M^+$, $M^{3+}$, $X^{VII}$) combinations are strongly desired to seek for the best candidates for photovoltaic applications.

Here we present via systematic first-principles calculations a study of phase stability and photovoltaic related properties for this class of inorganic Pb-free $A_2M^+M^{3+}X^{VII}_6$ halide perovskites. Our goal is to identify new stable $A_2M^+M^{3+}X^{VII}_6$ perovskites with potentially superior photovoltaic performance. Different from the previous theoretical work that focused on the $A_2M^+M^{3+}X^{VII}_6$ with $M^+$ limited to be group IB elements,[61,63,64] we consider a much broader range of monovalent elements for $M^+$ to accommodate different combinations of elements for the other sites. Specifically, we have considered combinations of (A, $M^+$, $M^{3+}$, $X^{VII}$) with $A = Cs^+$, $M^+$ = group IA ($Na^+$, $K^+$, $Rb^+$)/group IB ($Cu^+$, $Ag^+$, $Au^+$)/group IIIA ($In^+$, $Tl^+$), $M^{3+} = Bi^{3+}/Sb^{3+}$, and $X = F^-/Cl^-/Br^-/I^-$. As summarized in Figure 1a, in total there are 64 candidate compounds considered, of which only $Cs_2AgBiCl_6$, $Cs_2AgBiBr_6$, $Cs_2NaBiCl_6$, and $Cs_2KBiCl_6$ were experimentally synthesized.[56–64] Our results indicate that most of the materials in this family have phase stability against decomposition and show flexible tunability of optoelectronic properties with band gaps in the range from infrared to ultraviolet. Through photovoltaic-functionality-directed materials screening, we have identified 11 optimal compounds as promising Pb-free photovoltaic absorbers, as depicted in Figure 1b. Especially, we find two $In^+$ based compounds, $Cs_2InSbCl_6$ and $Cs_2InBiCl_6$, have direct optical band gaps of 1.02 and 0.91eV, respectively and they show high theoretical solar cell efficiencies comparable to that of $CH_3NH_3PbI_3$. Equally important, we have established the chemical trends of phase stability and optoelectronic properties for this novel class of $A_2M^+M^{3+}X^{VII}_6$ perovskites. Our work offers useful guidance for selectively utilizing these stable and Pb-free halide perovskites as promising solar absorbers.

## 2. Computational Methods

Our first-principles calculations were performed within the framework of density-functional theory (DFT) using the plane-wave pseudopotential approach as implemented in the VASP code.[65,66] The electron-core interactions are described with the frozen-core projected augmented wave pseudopotentials.[67] We use the generalized gradient approximation formulated by Perdew, Burke, and Ernzerhof (PBE)[68] as the exchange-correlation functional. The equilibrium structural parameters (including both lattice parameters and internal coordinates) of each candidate material are obtained by total energy minimization via the conjugate-gradient algorithm. The kinetic energy cutoffs for the plane-wave basis set are optimized to ensure the residual forces on atoms converged to below 0.0002 eV/Å. The k-point meshes with grid spacing of $2\pi\times0.025$ Å$^{-1}$ or less is used for electronic Brillouin zone integration. The electronic structures and optical absorption spectra are calculated by taking into account of the spin-orbit coupling (SOC) effect, with the Heyd-Scuseria-Ernzerhof (HSE) hybrid functional[69] remedying the underestimation of band gaps in common DFT-PBE calculations. The validity of our methodology in the $A_2M^+M^{3+}X^{VII}_6$ perovskite system is supported by good agreements on lattice parameters and band gaps between theory and available experiments (Table S1 in the Supporting Information). The comparison of optical absorption spectra between theory and experiment for two existing candidate $A_2M^+M^{3+}X^{VII}_6$ perovskites, $Cs_2AgBiCl_6$ and $Cs_2AgBiBr_6$ (Figure S1 in the Supporting Information) also indicates reasonably well agreement on absorption edges and relative intensities. Harmonic phonon spectrum is calculated with a finite-difference supercell approach,[70] and room-temperature phonon spectrum is obtained by taking into account anharmonic phonon-phonon interaction with a self-consistent *ab initio* lattice dynamical method.[71] To give an evaluation of photovoltaic performance of the selected optimal materials, the theoretical maximum solar cell efficiency, *i.e.*, "spectroscopic limited maximum efficiency (SLME)"[72,73] is calculated. Creation of calculation workflows, management of large amounts of calculations, extraction of calculated results, and post-processing analysis are performed by using an open-source Python framework designed for large-scale high-throughput energetic and property calculations, the Jilin University Materials-design Python Package (Jump[2], to be released soon). More details on calculations of photovoltaic-relevant properties are depicted in Supporting Information.

## 3. Results and Discussion

### 3.1 Phase stability of the class of $A_2M^+M^{3+}X^{VII}_6$ perovskite

We begin by probing the most energetically favored spatial distribution of $M^+X^{VII}_6$ and $M^{3+}X^{VII}_6$ motifs in $A_2M^+M^{3+}X^{VII}_6$ perovskites. We construct a 2×2×2 supercell of standard cubic perovskites, in which various types of arrangements of $M^+X^{VII}_6$ and $M^{3+}X^{VII}_6$ motifs are considered. In total, there are six different configurations. Figure 2a shows their calculated total energies for the case of $Cs_2AgBiCl_6$. Clearly the most stable configuration **F** is the standard double-perovskite structure (in space group of *Fm-3m*) with $M^+X^{VII}_6$ and $M^{3+}X^{VII}_6$ alternating along the three crystallographic axes and forming the rock-salt type ordering. This arrangement agrees with the structure type determined by the X-ray diffraction experiments.[61,74] The decrease of total energies from the configuration **A** to the **F** can be understood from the variation of electrostatic energies among $M^+$ and $M^{3+}$ cations, which show generally the same trend (as in Table S2 of Supporting Information). It is worth mentioning that other metastable configurations, especially **D** and **E**, may exist at finite temperatures, since their energies are only slightly higher (by less than 30 meV/atom than the ground-state **F**). We find that different arrangements of $M^+X^{VII}_6$ and $M^{3+}X^{VII}_6$ motifs can lead to a qualitative change of band gap feature, *i.e.*, from a zero gap in **A**, direct gaps in **B-E**, to an indirect gap in **F**. Besides, the gap values can also be widely modified quantitatively, *i.e.*, from 0 in **A** to 2.62 eV in **F**. These results imply that the possible disorder effect caused by varied arrangements of $M^+X^{VII}_6$ and $M^{3+}X^{VII}_6$ motifs at finite temperatures may offer an opportunity to further tune the optoelectronic properties of $A_2M^+M^{3+}X^{VII}_6$ perovskites. This phenomenon is worth further study. In the remaining, we focus on the most stable configuration — the ordered double-perovskite structure.

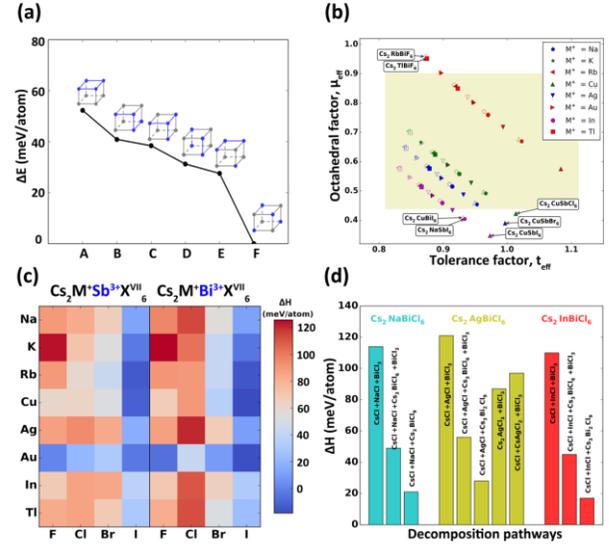

**Figure 2.** (a) Energies of $Cs_2AgBiCl_6$ with different types of $AgCl_6$ (in grey) +$BiCl_6$ (in blue) motifs arrangements. The energy of the lowest configuration **F** is set to zero. b) Distribution mapping of all the $A_2M^+M^{3+}X^{VII}_6$ perovskites with effective tolerance factor ($t_{eff}$) and octahedral factor ($\mu_{eff}$) as variables (red/green/blue/pink colors represent fluorides/chlorides/bromides/iodides; open/filled symbols correspond to Sb/Bi compounds). (c) Decomposition enthalpy ($\Delta H$) of $A_2M^+M^{3+}X^{VII}_6$ perovskites. (d) The $\Delta H$ corresponding to different decomposition pathways for selected $A_2M^+M^{3+}X^{VII}_6$.

To evaluate the crystallographic stability of materials in the perovskite structure, two empirical quantities in the framework of the idealized solid-sphere model, the Goldschmidt tolerance factor $t$ and the octahedral factor $\mu$, are usually considered. Previous statistic analysis of all the existing halide perovskites indicate that formability of perovskites requires $0.44 < \mu < 0.90$ and $0.81 < t < 1.11$.[75] In the current double-perovskite $A_2M^+M^{3+}X^{VII}_6$, one can define the effective $t_{eff}$ and $\mu_{eff}$ as follows:

$$t_{eff} = (R_A + R_{X^{VII}})/\sqrt{2}\{(R_{M^+} + R_{M^{3+}})/2 + R_{X^{VII}}\}, \quad (1)$$

and

$$\mu_{eff} = (R_{M^+} + R_{M^{3+}})/2R_{X^{VII}}, \quad (2)$$

where $R_A$, $R_{M^+}$, $R_{M^{3+}}$ and $R_{X^{VII}}$ are the ionic radii of A, $M^+$, $M^{3+}$ and $X^{VII}$ ions, respectively. Figure 2b shows the distribution of candidate $A_2M^+M^{3+}X^{VII}_6$ perovskites in the mapping with $t_{eff}$ and $\mu_{eff}$ as variables. We find that while most of $A_2M^+M^{3+}X^{VII}_6$ perovskites fall in the empirical stable area of perovskites (shaded), several materials including $Cs_2CuSbX^{VII}_6$ (with X=Cl, Bi and I), $Cs_2CuBiI_6$, $Cs_2NaSbI_6$, $Cs_2TlBiF_6$ and $Cs_2RbBiF_6$, are located outside. To evaluate thermodynamic stabilities of $A_2M^+M^{3+}X^{VII}_6$ perovskites, we calculate their decomposition energies with respect to possible decomposing pathways. The straightforward and predominate one is the decomposition of $A_2M^+M^{3+}X^{VII}_6$ into corresponding binary materials. Usually the halide perovskites are synthesized via its inverse reaction.[61–63,76] In particular, we calculate the decomposition enthalpy defined as

$$\Delta H = 2E[AX^{VII}] + E[M^+X^{VII}] + E[M^{3+}X^{VII}_3] - E[A_2M^+M^{3+}X^{VII}_6] \quad (3)$$

*i.e.*, the energy difference between the decomposed binary products and the $A_2M^+M^{3+}X^{VII}_6$ perovskite. Here, positive value of $\Delta H$ means energy gained from the decomposed products to the $A_2M^+M^{3+}X^{VII}_6$ perovskite, reflecting the thermodynamic stable condition of $A_2M^+M^{3+}X^{VII}_6$. Figure 2c shows our calculated $\Delta H$ of all the considered $A_2M^+M^{3+}X^{VII}_6$ perovskites. One observes that most of $A_2M^+M^{3+}X^{VII}_6$ demonstrate good stability with fairly large positive values of $\Delta H$ above 20 meV/atom. The exceptional cases with small and even negative $\Delta H$ occur in the iodides and the Au-contained materials. Clearly the values of $\Delta H$ diminish from fluorides/chlorides, bromides, to iodides. Further data analysis indicates a rough tendency of depressed $\Delta H$ with decreasing $t_{eff}$ and $\mu_{eff}$ (Figure S2 in the Supporting Information). We note that while certain materials (*e.g.*, $Cs_2AuBiF_6$ and $Cs_2KSbI_6$) within the crystallographically stable region (Figure 2b) are not stable with negative $\Delta H$, some materials (*e.g.*, $Cs_2CuSbBr_6$ and $Cs_2RbBiF_6$) outside of the crystallographically stable region are actually stable with positive $\Delta H$. This inconsistency between crystallographic stability and thermodynamic stability might be partially due to the mixed ionic and covalent bonding and multinary feature of $A_2M^+M^{3+}X^{VII}_6$ perovskites, making it difficult to assign realistic $t$ and $\mu$ for evaluating crystallographic stability. Besides the decomposition pathway into binaries, we also consider the possible decomposition processes involving ternary compounds. Figure 2d shows the results for $Cs_2M^+BiCl_6$ [$M^+$ = Na (group IA), Ag (IB) and In (IIIA)]. As seen, although the energy gains reflected by $\Delta H$ are reduced compared to the binary decomposition cases, the conclusions about their thermodynamic stabilities still hold. We note that four $A_2M^+M^{3+}X^{VII}_6$ perovskites that have been synthesized in experiments, *i.e.*, $Cs_2AgBiCl_6$, $Cs_2AgBiBr_6$, $Cs_2NaBiCl_6$, and $Cs_2KBiCl_6$,[56–64] all show robust thermodynamic stabilities based on the above analyses. This implies that other candidate $A_2M^+M^{3+}X^{VII}_6$ we predicted stable are promising to be synthesized in experiment.

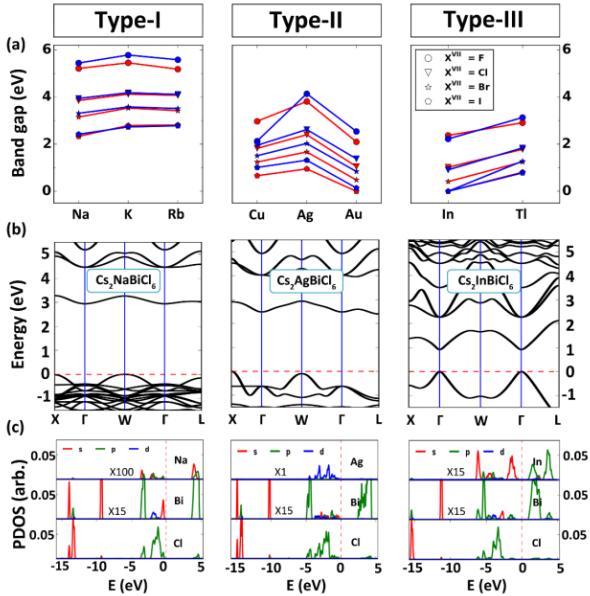

**Figure 3.** (a) Variation of band gaps with the M element for three categories of $A_2M^+M^{3+}X^{VII}_6$ perovskites (from left to right). The Bi/Sb compounds are shown in blue/red colors. (b, c) Electronic band structures and orbital-projected density of states of typical materials in the three categories: left panel ($Cs_2NaBiCl_6$, type-I), middle ($Cs_2AgBiCl_6$, type-II), right ($Cs_2InBiCl_6$, type-III).

**3.2 Chemical trends of electronic properties and classification of subtypes for $A_2M^+M^{3+}X^{VII}_6$ perovskites**

We next turn to discuss electronic properties of $A_2M^+M^{3+}X^{VII}_6$ perovskites. In terms of specific electron configuration of $M^+$, the only chemical specie spanning different elemental groups, this family of materials can be naturally classified into three categories: the type-I case with $M^+$ being the group IA elements ($Na^+$, $K^+$ and $Rb^+$) that have empty *s* and *d* outmost orbitals, the type-II with $M^+$ being the group IB elements ($Cu^+$, $Ag^+$ and $Au^+$) that have empty *s* but full *d* outmost orbitals, and the type-III with $M^+$ being the group IIIA elements ($In^+$ and $Tl^+$) that have full *s* outmost orbitals. Figure 3a shows band gap variation, and Figure 3b, 3c show electronic band structures and orbital-projected density of states of typical materials in the three categories, respectively. We can see with $X^{VII}$ changing from $F^-$, $Cl^-$, $Br^-$ to $I^-$, the band gaps show general decrease. For most of materials, as $M^{3+}$ changes from $Sb^{3+}$ to $Bi^{3+}$, the band gaps increase. These general trends on band gap can be understood in terms of analysis of constituents for band-edge states. It has been established that in $AM^{IV}X^{VII}_3$ perovskites the valence band maximum (VBM) is formed by anti-bonding states between $M^{IV}$-*s* and $X^{VII}$-*p* orbitals, and the conduction band minimum (CBM) mainly originates from $M^{IV}$-*p* orbitals.[24,25,55] For the $A_2M^+M^{3+}X^{VII}_6$ cases, the VBM is composed of anti-bonding states between $M^+/M^{3+}$-*s/d* orbitals and $X^{VII}$-*p* orbitals, and the CBM is mainly dominated by $M^+/M^{3+}$-*p* characters (Figure 3c, see also below). Walking from fluorides, chlorides, bromides, to iodides, the $X^{VII}$-*p* orbitals become higher in energy, resulting in raised VBM and thus reduced band gaps. Due to the Mass-Darwin relativistic effect, the Bi-*6s* orbital is lower in energy than the Sb-*5s* orbital, giving rise to the lower VBM and the larger band gaps in Bi-contained materials. This situation could reverse when $M^+$ belongs to group IB or group IIIA elements (Figure 3a), because in these systems the VBM may be no longer purely derived from the anti-bonding hybridization between Bi/Sb-*s* and $X^{VII}$-*p*, and the strong *p-d* coupling (for group IB elements) and *p-s* coupling (for group IIIA elements) play critical roles.

Due to the mix of $M^+$ and $M^{3+}$ ions at the octahedral site in double-perovskite structures, the quaternary $A_2M^+M^{3+}X^{VII}_6$ family shows more complicated electronic structures than those of the ternary $AM^{IV}X^{VII}_3$. We offer detailed discussion in terms of the three categories mentioned above as follows:

(i) *Type-I $A_2M^+M^{3+}X^{VII}_6$ with $M^+ = Na^+$, $K^+$ and $Rb^+$ (typical material: $Cs_2NaBiCl_6$)*. For this case, the VBM derives purely from the anti-bonding state between Bi-*s* and Cl-*p* (left of Figure 3c), resembling the $AM^{IV}X^{VII}_3$ perovskite case with the VBM formed by the anti-bonding of $M^{IV}$-*s* and $X^{VII}$-*p* orbitals. However, due to the reduced symmetry of double perovskites (*i.e.*, the *p-s* hybridization between Na and Cl being absent), the *p-s* coupling at the W and X points is larger than that at the Γ point. Consequently, the VBM is located at the W rather than Γ point (left of Figure 3b), giving a large indirect band gap of above 3.0 eV in $Cs_2NaBiCl_6$. This result is in accord with the reported value of hybrid double perovskite $(CH_3NH_3)_2KBiCl_6$.[77] The CBM state is a mix of Bi-*p* orbital, *s* orbital of ionic Na and Cl-*p* states. As $M^+$ changes from $Na^+$, $K^+$ to $Rb^+$, the $M^+$-*s* orbital energy increases, lifting the CBM.

Meanwhile, the increased size of $M^+$ from $Na^+$, $K^+$ to $Rb^+$ expands $M^+X^{VII}_6$ octahedra and shrinks $M^{3+}X^{VII}_6$ octahedra, resulting in enhanced p-s coupling between $M^{3+}$ and $X^{VII}$ and thus elevated VBM. Depending on specific amount of the up-shifts of both CBM and VBM states, the band gaps could be increased or decreased as $M^+$ changing from Na, K to Rb. This accounts for the slightly bowing-up profile of band gap variation in the left of Figure 3a.

(ii) *Type-II $A_2M^+M^{3+}X^{VII}_6$ with $M^+ = Cu^+$, $Ag^+$ and $Au^+$ (typical material: $Cs_2AgBiCl_6$)*. Here while the VBM is mainly dominated by the anti-bonding hybridized state of Ag-d and Cl-p orbitals, it is also contributed by the Bi-s state in similar binding energy (middle of Figure 3c). The electronic symmetry reduction due to the orbital mismatch between Ag-d and Bi-s leads to a larger off-centered p-d coupling. As the result the VBM moves from the Γ point to the X point, making $Cs_2AgBiCl_6$ an indirect-band-gap material. Similar to the type-I case, the CBM of $Cs_2AgBiCl_6$ originates mainly from a mix of Bi-p orbital, Ag-s and Cl-p states. Differently, as $M^+$ changes from $Cu^+$, $Ag^+$, to $Au^+$, the band gap variation is dominated by the change of the VBM states. Because Ag has the lowest d orbital among these three group IB elements,[78] the associated p-d repulsion is the weakest in Ag-contained compounds, leading to their relatively low VBM states and the largest band gaps (middle of Figure 3a).

(iii) *Type-III $A_2M^+M^{3+}X^{VII}_6$ with $M^+ = In^+$ and $Tl^+$ (typical material: $Cs_2InBiCl_6$)*. Because both $In^+$ and $Bi^{3+}$ ions have the same fully occupied outmost s shells similar to $Pb^{2+}$, the band edge structure of $Cs_2InBiCl_6$ (right of Figure 3b) is quite similar to those of $APbX^{VII}_3$.[24,25,55] The VBM is located at the zone-center Γ point because of absence of the electronic symmetry reduction in the above type-I and II cases. The band gap is thus direct. The VBM derives predominately from the anti-bonding state of In-s and Cl-p orbitals due to the higher energy of In-s orbital than that of Bi-s orbital, and the CBM is composed of Bi-p, In-p and Cl-p states (right of Figure 3c). Both the VBM and the CBM (located at Γ) have large band dispersion. This corresponds to quite small carrier effective masses (0.39 and 0.17 $m_0$ for electron and hole, respectively, according to our calculations) and strongly delocalized band-edge wave-functions, which are good for carrier extraction. With $M^+$ moving from In to Tl, because of the lower energy of Tl-6s than In-5s and the larger atomic size of Tl than In, the s-p coupling in Tl-contained systems is largely reduced compared to that in In-contained systems. As a result, the VBMs are lowered and the band gaps of Tl-contained systems show substantial increase (Figure 3a).

### 3.3 Combinatory materials screening of potentially high-photovoltaic-performance perovskites

With the data of phase stability and electronic property for all the considered $A_2M^+M^{3+}X^{VII}_6$ perovskites in hand, we resort to combinatory materials screening process to identify the potentially superior solar absorbers by considering the general requirements of photovoltaic functionality. This is done in terms of the following four criterions: (i) good thermodynamic stability — the positive decomposition enthalpy, (ii) high light absorption efficiency — the band gap sitting in the range of 0.8-2.0 eV, (iii) beneficial for ambipolar carriers conduction and efficient carriers extraction — both electron and hole effective masses smaller than 1.0 $m_0$, (iv) fast photon-induced carrier dissociation — small exciton binding energy (< 100 meV). Our thorough screening process is depicted in Figure 1b with the explicit data summarized in Table S3 of Supporting Information. As demonstrated we have identified 14 winning compounds, including nine type-II $A_2M^+M^{3+}X^{VII}_6$ (*i.e.*, $Cs_2CuSbCl_6$, $Cs_2CuSbBr_6$, $Cs_2CuBiBr_6$, $Cs_2AgSbBr_6$, $Cs_2AgSbI_6$, $Cs_2AgBiI_6$, $Cs_2AuSbCl_6$, $Cs_2AuBiCl_6$, and $Cs_2AuBiBr_6$) and five type-III ones (*i.e.*, $Cs_2InSbCl_6$, $Cs_2InBiCl_6$, $Cs_2TlSbBr_6$, $Cs_2TlSbI_6$, and $Cs_2TlBiBr_6$). None of them were previously synthesized in experiments. Considering the toxicity of Tl, we further exclude three Tl-contained compounds and have 11 optimal compounds left (as depicted in the last line of Figure 1b).

Among the eleven optimal materials, nine type-II $A_2M^+M^{3+}X^{VII}_6$ have indirect band gaps. Under normal circumstances such indirect-gap materials correspond to low light absorption efficiency and thus require relatively thick layers to absorb sufficient solar energy for photovoltaic applications. In this regard, we find that three materials, $Cs_2AgSbI_6$, $Cs_2AgBiI_6$, and $Cs_2AuBiBr_6$, with appropriate direct band gaps of 1.68, 1.77 and 1.83 eV, respectively (indirect gaps of 0.95, 1.32 and 0.84 eV, respectively), may deserve further investigation, since their visible light absorption efficiencies without including the phonon-assisted contributions seem not bad according to our calculations (Figure 4a). Note that the indirect-gap features of solar absorbers are known to be beneficial for low carrier recombination rates and thus long minority carrier diffusion lengths.

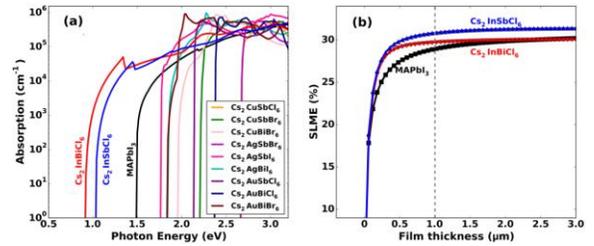

**Figure 4.** (a) Calculated absorption spectra of selected optimal $A_2M^+M^{3+}X^{VII}_6$ perovskites passing the screening process. (b) Simulated theoretical maximum solar cell efficiency, *i.e.*, "spectroscopic limited maximum efficiency" (SLME) of $Cs_2InSbCl_6$ and $Cs_2InBiCl_6$ as functions of film thicknesses. The results of $CH_3NH_3PbI_3$ are shown for comparison.

Two type-III optimal $A_2M^+M^{3+}X^{VII}_6$ perovskites, $Cs_2InSbCl_6$ and $Cs_2InBiCl_6$ have the band gap values of 1.02 and 0.91 eV, similar to that of silicon (1.14 eV), but in the direct nature, making them ideal candidates as high-performance solar absorbers. As shown in Figure 4a, they exhibit large intensity of band-edge optical transitions. This can be attributed to the strong intra-atomic In-s to In-p transition as well as inter-atomic Cl-p to In-p and Sb/Bi-p transitions with high joint density of states like the situation in the $APbX^{VII}_3$ case.[24] Figure 4b shows our calculated SLME[72,73] values of them, as the evaluation of their photovoltaic performance. One clearly sees that the SLMEs of $Cs_2InSbCl_6$ and $Cs_2InBiCl_6$ are comparable to (or even higher than) that of $CH_3NH_3PbI_3$. Particularly at 1 μm film thickness, their SLME values reach 31% and 30%, slightly surpassing 29% of $CH_3NH_3PbI_3$.

We further examine the dynamic phonon stability for two best-of-class compounds $Cs_2InSbCl_6$ and $Cs_2InBiCl_6$. The results are shown in Figure S3 of Supporting Information. We find that the harmonic phonon spectra of $Cs_2InBiCl_6$ and $Cs_2InSbCl_6$ (at 0 K) show substantial imaginary optical modes (black lines in Figure S3a and S3b). However when we take into account phonon-phonon interaction (anharmonic effect) at room temperature (300 K),[71] all the imaginary phonons disappear (red lines in Figure S3a and S3b). The results indicate that $Cs_2InBiCl_6$ and $Cs_2InSbCl_6$ are dynamically stable under ambient condition. This conclusion is supported by the fact that the experimentally synthesized compound, $Cs_2AgBiCl_6$, show the similar behavior, *i.e.*, unstable harmonic phonons being stabilized by room-temperature anharmonic interaction (Figure S3c). Here phonon entropy at finite temperatures plays critical role in phonon spectrum renormalization. We note that though the monovalent $In^+$ ions are less common than the trivalent $In^{3+}$ ions in materials, the $In^+$-contained compounds do exist substantially.[79–81] In fact, InCl is a directly available precursor for synthesizing the above two compounds. To probe whether the oxidation of $In^+$ into $In^{3+}$ can cause material instability, we consider the possible decompositions of $Cs_2InSbCl_6$ and $Cs_2InBiCl_6$ into their corresponding $In^{3+}$-contained compounds (Figure S4 of Supporting Information). The safely large positive $\Delta H$ values (above 100 meV/atom) indicate the maintained stabilities of $Cs_2InSbCl_6$ and $Cs_2InBiCl_6$ against decompositions into their In-related oxidized phases.

As known, a dominant advantage of $CH_3NH_3PbI_3$ as high-efficiency solar absorber is the defect-tolerant feature, i.e., almost all the low-formation-energy defects are shallow rather than deep states.[25–29] This is responsible for its bipolar conducting property and long carrier diffusion length (owing to absence of charge trapping centers). Because the $M^{3+}$ ion in $A_2M^+M^{3+}X^{VII}_6$ adopts the $Sb^{3+}/Bi^{3+}$ that have the same electron configuration ($ns^2$) with $Sn^{2+}/Pb^{2+}$, it is reasonably conjectured that similar defect-tolerant feature might be to some extent at play here. Taking $Cs_2InSbCl_6$ and $Cs_2InBiCl_6$ as instances, the easily formed p-type defects $V_{Sb/Bi}$ and $V_{In}$ are expected to be shallow by consideration of the anti-bonding (between $In^+/Bi^{3+}$-s and Cl-p orbitals) feature of valence bands; the most likely n-type defects $Cs_i$ and $V_{Cl}$ might be shallow as well due to rather strong ionic bonding of this family of materials.[25]

Finally, we note that two best-of-class compounds $Cs_2InSbCl_6$ and $Cs_2InBiCl_6$ are distinct in structure, electronic and optoelectronic properties from other Bi-containing perovskites such as $MA_3Bi_2I_9$ and $Cs_3Bi_2I_9$[82]. $MA_3Bi_2I_9/Cs_3Bi_2I_9$ has a crystal structure composed of isolated face-shared $BiI_6$ octahedra, different with the standard perovskite structure of $MAPbI_3$ formed in three-dimensional corner-shared $PbI_6$ octahedra framework. As the result, they have an oversized, indirect band gap of more than 2.0 eV, giving low conversion efficiencies merely up to ~1%.[82] In contrast, $Cs_2InBiCl_6$ and $Cs_2InSbCl_6$ have quite similar electronic band structures as $MAPbI_3$, because of the same underlying perovskite structure, the isoelectronic feature and the octahedral B site occupied by the cations all with lone pair s states. This is clearly indicated by the direct comparison of band structure between $Cs_2InBiCl_6/Cs_2InSbCl_6$ and $Cs_2Pb_2Cl_6$ (an equivalent of MAPbI$_3$) as shown in Figure S5 of Supporting Information. The electronic and optoelectronic properties of $Cs_2InBiCl_6$ and $Cs_2InSbCl_6$ resembling those of $MAPbI_3$ promises their high solar cell conversion efficiencies.

## 4. Conclusions

In summary, we have exploited the idea of cation-transmutation to rationally design stable Pb-free halide perovskites for photovoltaic applications. By transmuting two divalent Pb ions in $APbX^{VII}_3$ perovskites into one monovalent ion $M^+$ and one trivalent ion $M^{3+}$, a rich class of quaternary $A_2M^+M^{3+}X^{VII}_6$ materials in double-perovskite structure can be formed. By using systematic first-principles calculations, we find this class of materials generally shows good phase stability against decomposition and has diverse electronic properties with wide-range tunable band gaps. Classification of subtypes in terms of electronic structure feature is proposed and the chemical trends of phase stability and electronic properties of each subtype are established. Photovoltaic-functionality-directed material screening process involving totally 64 candidate materials allows us to identify 11 non-toxic $A_2M^+M^{3+}X^{VII}_6$ perovskites as promising absorbers to replace $APbX^{VII}_3$ in halide perovskite solar cells. They show intrinsic thermodynamic stabilities, suitable band gaps, small carrier effective masses, low exciton binding energies, etc. Among them, two direct-gap materials, $Cs_2InSbCl_6$ and $Cs_2InBiCl_6$, with the gap values around 1.0 eV, show the theoretical maximum solar cell efficiencies comparable to that of $CH_3NH_3PbI_3$. Halide perovskites containing the alternative ions with $ns^2np^0$ configuration (other than Pb) such as Bi have shown potential as solar absorbers, but so far exhibiting rather low performance.[82,83] Our work offers a promising routine for development of halide perovskite absorbers with the $ns^2np^0$ ions to eliminate toxic Pb in perovskite solar cells. Experimental efforts to synthesize our designed materials with potentially superior photovoltaic performance are strongly called for.

## Supporting Information

More detailed computational procedures, comparison between calculated and experimental data for existing $A_2M^+M^{3+}X^{VII}_6$ perovskites, additional supporting photovoltaic-related data on materials stability analysis, explicit calculated data of all the candidate $A_2M^+M^{3+}X^{VII}_6$ used for materials screening, phonon spectra of best-of-class winning compounds.

## Corresponding Author

* suhuaiwei@csrc.ac.cn
* lijun_zhang@jlu.edu.cn

## Author Contributions

‡These authors contributed equally.

## Acknowledgement

The authors acknowledge stimulating discussion with Dr. B Chen, and funding support from the Recruitment Program of Global Youth Experts in China, National Key Research and Development Program of China (under Grants No. 2016YFB0201204), and National Natural Science Foundation of China (under Grant No. 11404131). Work at Beijing CSRC is supported by NSFC under Grant Number U1530401 and National Key Research and Development Program of China under Grant No. 2016YFB0700700. Part of calculations was performed in the high performance computing center of Jilin University and on TianHe-1 (A) of the National Supercomputer Center in Tianjin.

# Supporting Information

# Design of Lead-free Inorganic Halide Perovskites for Solar Cells via Cation-transmutation

Xin-Gang Zhao[1,‡], Ji-Hui Yang[2,‡], Yuhao Fu[1], Dongwen Yang[1], Qiaoling Xu[1], Liping Yu[3], Su-Huai Wei[4,*] and Lijun Zhang[1,*]

[1]Department of Materials Science and Engineering and Key Laboratory of Automobile Materials of MOE and State Key Laboratory of Superhard Materials, Jilin University, Changchun 130012, China

Jilin University, Changchun 130012, China

[2]Department of Materials Science and Nanoengineering, Rice University, Houston, Texas 77005, USA

[3]Department of Physics, Temple University, Philadelphia, PA 19122, USA

[4]Beijing Computational Science Research Center, Beijing 100094, China

**Corresponding Author**

* suhuaiwei@csrc.ac.cn
* lijun_zhang@jlu.edu.cn

**Author Contributions**

[‡]X.-G.Z. and J.-H.Y. contributed equally.

# Computational Details

*Band Gap:* The DFT calculations are known to seriously underestimate the electronic band gap that is a critical quantity determining photovoltaic performance of materials. To remedy this problem, we employ the HSE hybrid functional containing the standard 25% of the exact Fock exchange to reduce the self-interaction error and reach correct gap values. The SOC effect, which plays critical role in determining electronic structures of the compounds containing heavy elements such as Bi/Sb, is included. Considering the heavy computational cost of HSE+SOC calculations for a large number of candidate compounds for materials screening, we adopt the following compromised approach: firstly we use the PBE+SOC calculations (in dense enough *k*-point meshes, with grid spacing of $2\pi \times 0.01$ Å$^{-1}$ or less) to determine the *k*-points at which the band-edge states lie; then such *k*-points are passed to the HSE+SOC calculations (in moderate *k*-point meshes, with grid spacing of about $2\pi \times 0.03$ Å$^{-1}$), from which the gap values are obtained. The assumption here is the HSE calculations do not make change to the band structure shape from the DFT-PBE calculations, which has been demonstrated to be the case in most of materials.[1,2] The good agreements of gap values between theory and experiment for $Cs_2AgBiCl_6$ and $Cs_2AgBiBr_6$ (Table S1 in the Supporting Information) further validate the above approach. After obtaining the band gaps from the HSE+SOC, the band structure, (projected) density of states, and absorption spectrum from the PBE+SOC calculations are corrected by the scissor operator to match the corresponding HSE+SOC gap values.

*Carriers Effective Masses:* We employ the semi-classical Boltzmann transport theory[3] to process band structure for getting the effective mass tensors that relate directly to electrical conductivity. In this way the effects of band non-parabolicity, anisotropy of bands, multiple bands coupling, etc. on carrier transport are synthetically taken into account. We perform the PBE+SOC calculations at the more dense *k*-points grid (of $2\pi \times 0.005$ Å$^{-1}$) to guarantee the convergence of such transport related calculations. The room-temperature mass values corresponding to the carrier concentration of $1.0 \times 10^{18}$ cm$^{-3}$ are taken.

*Exciton binding energy:* To probe the feasibility of photon-induced exciton dissociation in solar cells, we calculate the exciton binding energy by adopting the hydrogen-like Wannier-Mott exciton model. The essential input parameters are $m^*$ of electron and hole, and dielectric constant. Particularly, the $E_b$ is given by $E_b = \mu^* R_y / m_0 \varepsilon_r^2$, where $\mu^*$ is the reduced exciton mass (i.e. $1/\mu^* = 1/m_e + 1/m_h$), $R_y$ is the atomic Rydberg energy, and $\varepsilon_r$ is the relative dielectric constant. The high-frequency limit of dielectric constant ($\varepsilon_\infty$) caused by electronic polarization, is taken as $\varepsilon_r$; the resulted $E_b$ describes the excitons generated immediately after photon excitation (without lattice polarization process involved).

*Phonon spectrum:* The harmonic phonon spectrum is calculated from second-order interatomic force constants obtained by using the real-space finite-difference supercell approach implemented in Phonopy code.[4] We take the $2 \times 2 \times 2$ supercell of the primitive lattice of the double-perovskite *Fm-3m* structure. The *k*-point mesh with grid spacing of $2\pi \times 0.03$ Å$^{-1}$ is used for electronic Brillouin zone integration. The room-temperature phonon spectrum is obtained by taking into account anharmonic phonon-phonon interaction with a self-consistent *ab initio* lattice dynamical (SCAILD) method.[5] This is done via calculating the phonon frequencies renormalization induced by phonon entropy, *i.e.*, the geometric disorder introduced by several frozen phonons simultaneously presenting in the simulated supercell. The SCAILD method alternates between creating atomic displacements in terms of phonon frequencies and evaluating phonon frequencies from calculated forces acting on the displaced atoms. The self-consistent cycle was terminated when the difference in the approximate free energy between two consecutive iterations is less than 1 meV. Calculations are performed at constant volume with thermal expansion effect ignored.

*Absorption spectrum:* The photon energy ($\omega$) dependent absorption coefficient $\alpha(\omega)$ is calculated in terms of the following relation,

$$a(w) = \frac{\sqrt{2}\varepsilon_2(w)w}{c\sqrt{\varepsilon_1(w) \pm \sqrt{\varepsilon_1(w)^2 + \varepsilon_2(w)^2}}},$$

where $\varepsilon_1(\omega)/\varepsilon_2(\omega)$ are real/imaginary parts of dielectric function, and c is speed of light. The $\varepsilon_2(\omega)$ is calculated in the random phase approximation,[6] and $\varepsilon_1(\omega)$ is evaluated from $\varepsilon_2(\omega)$ via the Kramers-Kronig relation. The dense k-point meshes with grid spacing of $2\pi \times 0.015$ Å$^{-1}$ or less is used for calculating conduction and valence band states to ensure $\varepsilon_2(\omega)$ converged. The number of empty conduction band states used for such calculations is twice of the number of valence bands.

*Spectroscopic limited maximum efficiency:* The maximum solar cell efficiency is simulated through calculating spectroscopic limited maximum efficiency (SLME) based on the improved Shockley-Queisser model. The SLME of a material takes into account the band gap size, the band gap type (direct versus indirect), and the optical absorption spectrum, all of which can be obtained from reliable first principles calculations. The calculation of radiative and non-radiative recombination current is based on detailed balance theory using the energy difference between the minimum band gap and direct-allowed gap as the input. The detailed calculation procedure was described elsewhere.[7,8] The simulation is performed under the standard AM1.5G solar spectrum at room temperature.

**Table S1:** Comparison between theory and experiment for cubic lattice parameters ($a$) and band gaps ($E_g$) of the $A_2M^+M^{3+}X^{VII}_6$ perovskites synthesized experimentally. One clearly sees good agreement between theoretical and experimental data. This validates the reliability of first-principles computational approach used for our materials screening.

|  | $a$ (Å) | | $E_g$ (eV) | |
| --- | --- | --- | --- | --- |
|  | Exp. | Cal. | Exp. | Cal. |
| $Cs_2NaBiCl_6$ | 10.839[a] | 11.023 | -- | 3.96 |
| $Cs_2AgBiCl_6$ | 10.777[b] | 10.944 | 2.77[b]/2.2[d] | 2.62 |
| $Cs_2AgBiBr_6$ | 11.250[b]/11.271[c] | 11.473 | 1.95[b]/2.19[c]/1.9[d] | 2.03 |

a). Smit W M A, Dirksen G J, Stufkens D J. *Journal of Physics and Chemistry of Solids,* **1990**, 51(2): 189-196.

b). McClure E T, Ball M R, Windl W, et al. *Chem. Mater.*, **2016**, 28(5): 1348-1354.

c). Slavney A H, Hu T, Lindenberg A M, et al. *J. Am. Chem. Soc.*, **2016**, 138(7): 2138-2141.

d). Filip M R, Hillman S, Haghighirad A A, et al. *J. Phys. Chem. Lett.*, **2016**, 7(13): 2579-2585.

**Table S2:** Calculated decomposition entropy ($\Delta H$), electrostatic energies among $Ag^+$ and $Bi^{3+}$ cations ($\Delta E_{static}$), band gap ($E_g$), and band gap feature (direct/indirect) for the structural configurations with different spatial distribution of $AgCl_6$ and $BiCl_6$ motifs in the representative candidate $A_2M^+M^{3+}X^{VII}_6$ perovskite, $Cs_2AgBiCl_6$ (see Figure 2a and the main text). The values of $\Delta E_{static}$ are relative to that of the most stable configuration **F**.

| Quantities | Structural configurations | | | | | |
|---|---|---|---|---|---|---|
| | **A** | **B** | **C** | **D** | **E** | **F** |
| $\Delta H$ (meV/atom) | 69 | 80 | 83 | 90 | 93 | 120 |
| $\Delta E_{static}$ (a.u.) | 0.185 | 0.097 | 0.095 | 0.025 | 0.044 | 0.000 |
| $E_g$ (eV) | -- | 0.43 | 0.42 | 1.05 | 0.81 | 2.62 |
| Direct/Indirect gap (D/I) | -- | D | D | D | D | I |

**Table S3:** Explicit calculated data of cubic lattice constant (α), decomposition entropies (ΔH), band gaps ($E_g$), electron and hole effective masses ($m_e$, $m_h$), and exciton binding energies ($\Delta E_b$) of all the candidate $A_2M^+M^{3+}X^{VII}_6$ perovskites for materials screening. The optimal winning compounds satisfying all the criteria of the screening (the last line of Figure 1b) are shaded by green. The calculated data with the same approach for the prototype material, MAPbI$_3$, are shown for comparison.

| $M^+$ | $X^{VII}$ | $Cs_2MSbX^{VII}_6$ | | | | | | $Cs_2MBiX^{VII}_6$ | | | | | |
|---|---|---|---|---|---|---|---|---|---|---|---|---|---|
| | | α (Å) | ΔH (meV/atom) | $E_g$ (eV) | $m_e$ ($m_0$) | $m_h$ ($m_0$) | $\Delta E_b$ (meV) | α (Å) | ΔH (meV/atom) | $E_g$ (eV) | $m_e$ ($m_0$) | $m_h$ ($m_0$) | $\Delta E_b$ (meV) |
| MAPbI$_3$ | | 6.32 | 2 | 1.49 | 0.18 | 0.42 | 52 | | -- | -- | -- | -- | -- |
| Na | F | 9.19 | 89 | 5.21 | -3.60 | 6.59 | 5151 | 9.28 | 95 | 5.52 | -3.60 | 6.59 | 5434 |
| Na | Cl | 10.93 | 83 | 3.85 | -0.65 | 3.16 | 782 | 11.02 | 114 | 4.07 | -0.65 | 3.16 | 839 |
| Na | Br | 11.51 | 67 | 3.15 | -0.47 | 2.78 | 443 | 11.61 | 55 | 3.38 | -0.47 | 2.78 | 479 |
| Na | I | 12.39 | 13 | 2.33 | -0.35 | 1.58 | 214 | 12.49 | 12 | 2.42 | -0.35 | 1.58 | 236 |
| K | F | 9.63 | 124 | 5.45 | -2.29 | 4.46 | 3616 | 9.74 | 126 | 5.78 | -2.29 | 4.46 | 3861 |
| K | Cl | 11.38 | 72 | 4.12 | -0.86 | 4.18 | 1204 | 11.48 | 102 | 4.18 | -0.86 | 4.18 | 1285 |
| K | Br | 11.97 | 53 | 3.53 | -0.62 | 4.55 | 733 | 12.07 | 42 | 3.58 | -0.62 | 4.55 | 787 |
| K | I | 12.83 | -3 | 2.79 | -0.47 | 1.98 | 374 | 12.93 | -4 | 2.74 | -0.47 | 1.98 | 403 |
| Rb | F | 9.83 | 90 | 5.18 | -1.85 | 2.52 | 2580 | 9.96 | 91 | 5.58 | -1.85 | 2.52 | 2775 |
| Rb | Cl | 11.61 | 60 | 4.07 | -0.87 | 2.22 | 1099 | 11.72 | 90 | 4.11 | -0.87 | 2.22 | 1178 |
| Rb | Br | 12.15 | 44 | 3.42 | -0.64 | 2.30 | 664 | 12.28 | 42 | 3.51 | -0.64 | 2.30 | 755 |
| Rb | I | 13.04 | -14 | 2.81 | -0.49 | 2.27 | 424 | 13.14 | -15 | 2.78 | -0.49 | 2.27 | 457 |
| Cu | F | 9.10 | 61 | 2.97 | -0.79 | 2.11 | 533 | 9.18 | 66 | 2.13 | -0.79 | 2.11 | 515 |
| Cu | Cl | 10.52 | 63 | 1.82 | -0.38 | 0.66 | 97 | 10.61 | 90 | 1.95 | -0.38 | 0.66 | 115 |
| Cu | Br | 11.07 | 52 | 1.24 | -0.29 | 0.53 | 49 | 11.17 | 36 | 1.51 | -0.29 | 0.53 | 61 |
| Cu | I | 11.87 | -11 | 0.66 | -0.21 | 0.45 | 20 | 11.97 | -18 | 1.01 | -0.21 | 0.45 | 26 |
| Ag | F | 9.41 | 87 | 3.81 | -0.75 | 0.98 | 645 | 9.52 | 87 | 4.14 | -0.75 | 0.98 | 733 |
| Ag | Cl | 10.84 | 95 | 2.40 | -0.38 | 0.52 | 163 | 10.94 | 121 | 2.62 | -0.38 | 0.52 | 190 |
| Ag | Br | 11.37 | 83 | 1.67 | -0.29 | 0.47 | 86 | 11.47 | 69 | 2.03 | -0.29 | 0.47 | 104 |
| Ag | I | 12.13 | 17 | 0.95 | -0.22 | 0.45 | 38 | 12.24 | 11 | 1.32 | -0.22 | 0.45 | 48 |
| Au | F | 9.50 | 1 | 2.10 | -0.64 | 0.93 | 272 | 9.59 | -6 | 2.54 | -0.64 | 0.93 | 436 |
| Au | Cl | 10.83 | 21 | 1.05 | -0.28 | 0.39 | 54 | 10.94 | 43 | 1.38 | -0.28 | 0.39 | 74 |
| Au | Br | 11.32 | 31 | 0.48 | -0.21 | 0.30 | 22 | 11.42 | 9 | 0.84 | -0.21 | 0.30 | 31 |
| Au | I | 12.06 | -3 | 0.00 | -0.01 | 0.01 | 0 | 12.17 | -17 | 0.14 | -0.01 | 0.01 | 1 |
| In | F | 9.70 | 68 | 2.37 | -0.61 | 0.39 | 283 | 9.80 | 69 | 2.22 | -0.61 | 0.39 | 320 |
| In | Cl | 11.32 | 86 | 1.02 | -0.39 | 0.17 | 60 | 11.44 | 110 | 0.91 | -0.39 | 0.17 | 81 |
| In | Br | 11.80 | 84 | 0.41 | -0.30 | 0.11 | 14 | 11.93 | 65 | 0.00 | -0.30 | 0.11 | 31 |
| In | I | 12.55 | 37 | 0.00 | -0.23 | 0.08 | 0 | 12.69 | 32 | 0.00 | -0.23 | 0.08 | 11 |
| Tl | F | 9.77 | 83 | 2.91 | -0.61 | 0.43 | 388 | 9.88 | 82 | 3.13 | -0.61 | 0.43 | 444 |
| Tl | Cl | 11.44 | 86 | 1.78 | -0.38 | 0.26 | 139 | 11.56 | 112 | 1.84 | -0.38 | 0.26 | 167 |
| Tl | Br | 11.94 | 68 | 1.25 | -0.30 | 0.20 | 70 | 12.07 | 51 | 1.28 | -0.30 | 0.20 | 89 |
| Tl | I | 12.72 | 31 | 0.81 | -0.23 | 0.18 | 35 | 12.84 | 24 | 0.78 | -0.23 | 0.18 | 45 |

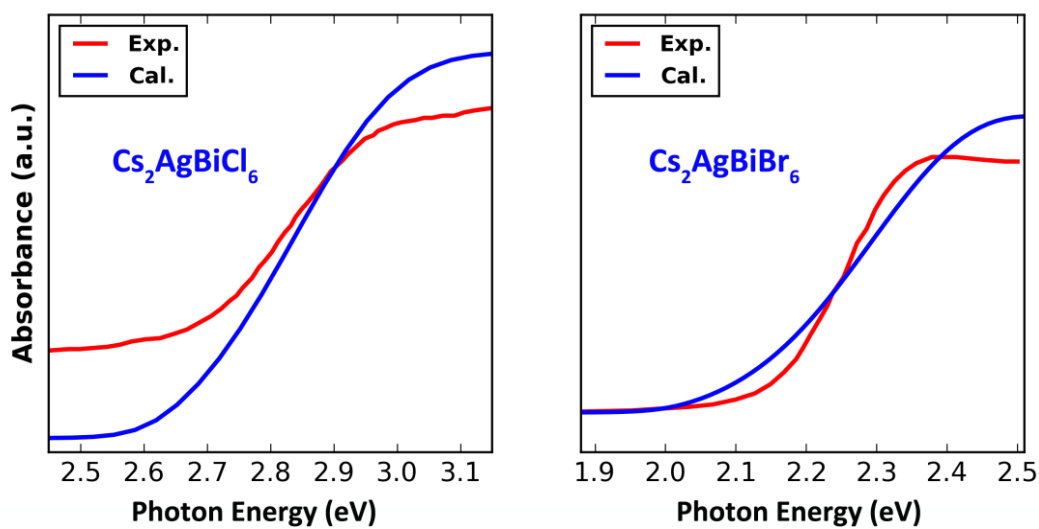

**Figure S1:** Comparison of optical absorption between theory and experiment for two existing candidate $A_2M^+M^{3+}X^{VII}_6$ perovskites, $Cs_2AgBiCl_6$ and $Cs_2AgBiBr_6$. The experimental data are taken from Ref. 64 (for $Cs_2AgBiCl_6$) and Ref. 63 (for $Cs_2AgBiBr_6$) of the main text.

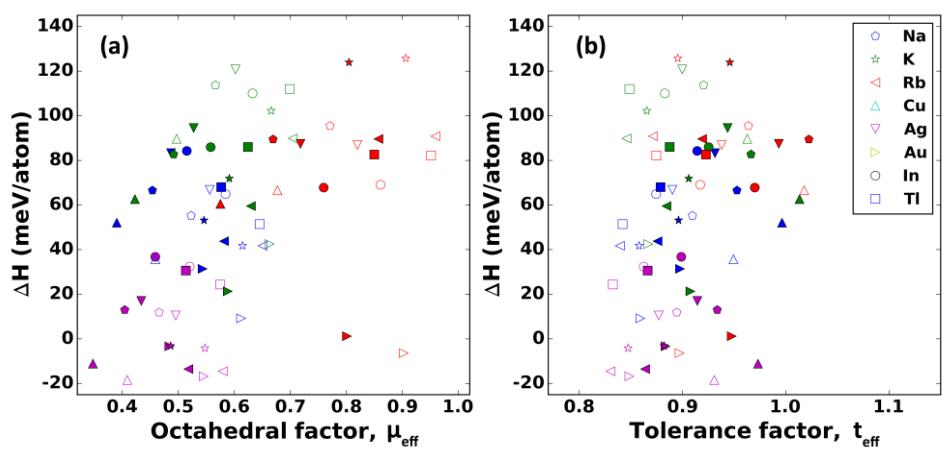

**Figure S2:** Decomposition entropy ($\Delta H$) as the function of the effective (a) tolerance factor ($t_{eff}$) and (b) octahedral factor ($\mu_{eff}$) for $A_2M^+M^{3+}X^{VII}_6$ perovskites. Red/green/blue/matron colors represent fluorides/chlorides/bromides/iodides, and open/filled symbols correspond to Sb/Bi compounds.

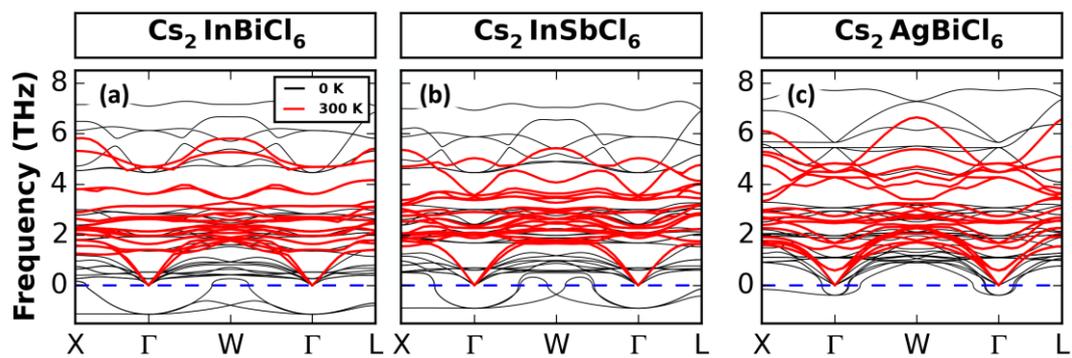

**Figure S3:** Calculated phonon spectra of $Cs_2InBiCl_6$ (a), $Cs_2InSbCl_6$ (b), and the experimentally synthesized $A_2M^+M^{3+}X^{VII}_6$ perovskite $Cs_2AgBiCl_6$ (c). The harmonic phonon spectra at 0 K are shown in black, and the room-temperature (300 K) phonon spectra are in red.

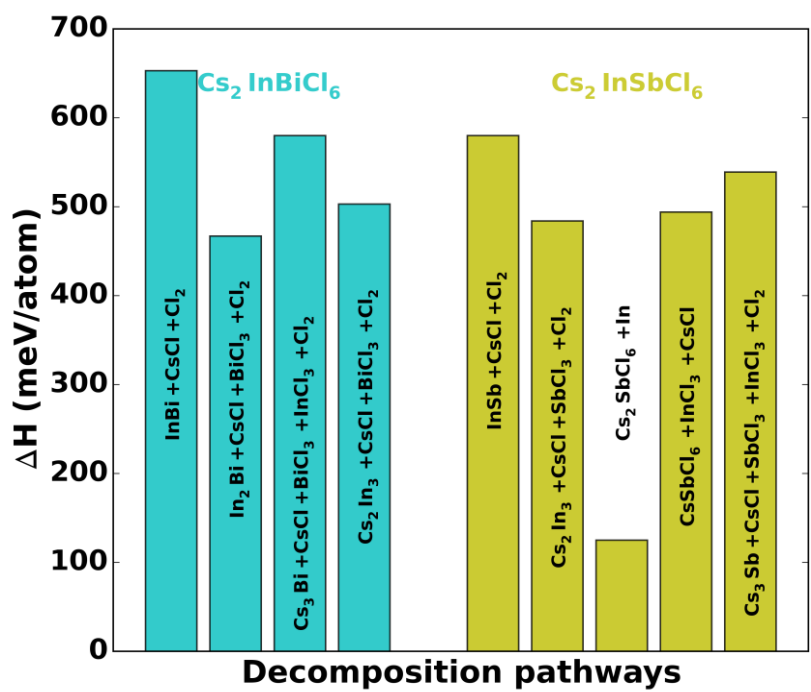

**Figure S4:** Decomposition entropy (Δ*H*) associated with the decomposition pathways involving In$^{3+}$-contained compounds for the potential high-photovoltaic-performance perovskites of Cs$_2$InSbCl$_6$ and Cs$_2$InBiCl$_6$.

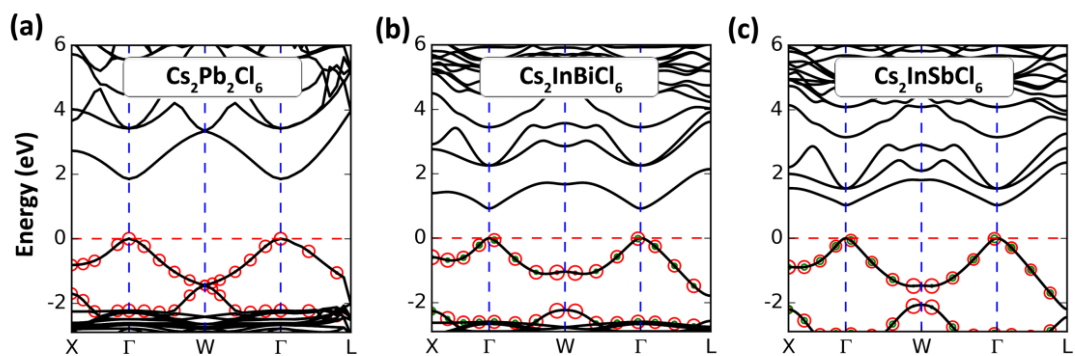

**Figure S5:** Comparison of calculated band structure among (a) "$Cs_2Pb_2Cl_6$" (*i.e.*, $CsPbCl_3$ represented in the double perovskite structure), (b) $Cs_2InBiCl_6$ and (c) $Cs_2InSbCl_6$. The orbital projection onto the lone pair *s* states is depicted with circles (in red for $Pb^{2+}$ and $In^+$, and green for $Sb^{3+}/Bi^{3+}$).